\documentclass[11pt]{article}
\usepackage{amssymb,amsmath,amsfonts}
\usepackage{graphicx}
\usepackage{graphics}
\usepackage{eepic,epsfig}

\begin{document}

\title{Synchrotron and Smith-Purcell radiations from a charge \\
rotating around a cylindrical grating}
\author{A. A. Saharian$^{1,2}$,\thinspace\, A. S. Kotanjyan$^{2}$%
,\thinspace\, A. R. Mkrtchyan$^{1,3}$,\thinspace\, B. V. Khachatryan$^{1,2}$
\\
%EndAName
\\
\textit{$^1$Institute of Applied Problems in Physics NAS RA,}\\
\textit{25 Nersessian Street, 0014 Yerevan, Armenia}\vspace{0.3cm}\\
\textit{$^2$Department of Physics, Yerevan State University,}\\
\textit{1 Alex Manoogian Street, 0025 Yerevan, Armenia}\vspace{0.3cm}\\
\textit{$^3$Tomsk Polytechnic University, 30 Lenin Avenue, 634050 Tomsk,
Russia}}
\maketitle

\begin{abstract}
We investigate the radiation from a charge rotating around conductors with
cylindrical symmetry. First the problem is considered with a charge rotating
around a conducting cylinder immersed in a homogeneous medium. The surface
charge and current densities induced on the cylinder surface are evaluated.
A formula is derived for the spectral-angular density of the radiation
intensity. In the second part, we study the radiation for a charge rotating
around a diffraction grating on a cylindrical surface with metallic strips
parallel to the cylinder axis. The effect of the grating on the radiation
intensity is approximated by the surface currents induced on the strips by
the field of the rotating charge. The expressions are derived for the
electric and magnetic fields and for the angular density of the radiation
intensity on a given harmonic. We show that the interference between the
synchrotron and Smith-Purcell radiations may lead to interesting features.
In particular, the behavior of the radiation intensity on large harmonics
can be essentially different from that for a charge rotating in the vacuum
or around a solid cylinder. Unlike to these limiting cases, for the geometry
of diffraction grating the radiation intensity on higher harmonics does not
vanish for small angles with respect to the cylinder axis. For given
characteristics of the charge, by the choice of the parameters of the
diffraction grating, one can have highly directional radiation near the
normal to the plane of the charge rotation. With decreasing energy, the
relative contribution of the synchrotron radiation decreases and the
Smith-Purcell part is dominant.
\end{abstract}

\bigskip

\section{Introduction}

Due to its unique characteristics, such as broad spectrum, high flux, and
high degree of polarization, the synchrotron radiation (for reviews see \cite%
{Soko86}-\cite{Will11}) is an ideal tool for many types of research and also
has found industrial applications including materials science, biological
and life sciences, medicine, chemistry. These extensive applications
motivate the importance of investigations for various types of mechanisms to
control the radiation parameters. In particular, it is of interest to
consider the influence of a medium on the spectral-angular characteristics
of the radiation intensity. The synchrotron radiation from a charged
particle circulating in a homogeneous and isotropic medium was considered in
\cite{Tsyt51,Kita60,Erbe76}, where it has been shown that the interference
between the synchrotron and Cherenkov radiations leads to remarkable
effects. In particular, for high-energy particles and close to the Cherenkov
threshold the total radiation rate oscillates.

New interesting features arise for the synchrotron radiation in
inhomogeneous media. As an exactly solvable problem of this kind, in \cite%
{Kota00}, we have studied the radiation from a charge orbiting around/inside
a dielectric cylinder immersed into a homogeneous medium. Under the
Cherenkov condition for the material of the cylinder and for the velocity of
the particle image on the cylinder surface, strong narrow peaks appear in
the angular distribution of the radiation intensity at large distances from
the cylinder. At these peaks the radiated energy exceeds the corresponding
quantity in the case of a homogeneous medium by several orders of magnitude.
It has been shown that the angular locations of the peaks are determined by
the equation that is obtained from the equation determining the eigenmodes
for the dielectric cylinder replacing the Hankel function of the first kind
by the Neumann function. Similar features for the radiation generated by a
charge moving along a helical orbit around/inside a dielectric cylinder have
been discussed in \cite{Saha05}. This type of electron motion is employed in
helical undulators.

Another type of the radiation that attracted a great deal of attention is
the Smith-Purcell radiation (for reviews see \cite{Shes98,Poty11}). It
arises when charged particles are in flight near a diffraction grating. The
Smith-Purcell radiation presents a tunable source of the electromagnetic
radiation over a wide range of frequency spectrum. It has a number of
remarkable properties and is widely used in various fields of science and
technology for the generation of electromagnetic radiation in different
wavelength ranges and for the determination of the characteristics of
emitting particles using the properties of the radiation field. In
particular, the Smith-Purcell radiation is one of the basic mechanisms for
the generation of electromagnetic waves in the millimeter and submillimeter
wavelengths. The radiation having similar physical properties and induced by
a charged particle in flight over a surface acoustic wave has been discussed
in \cite{Mkrt91} (for a more general problem of the radiation from a charge
intersecting a periodically deformed dynamical boundary between two media
see \cite{Mkrt16}). The use of surface waves simplifies the control of the
period and the amplitude of the periodic structure and, hence, the
angular-frequency characteristics of the emitted radiation. Compared to the
case of diffraction grating, the dynamical nature of the periodic structure
leads to an additional shift in the radiation frequency for a given
direction.

The previous investigations of the Smith-Purcell effect mainly consider the
radiation sources moving along straight trajectories. In this case the
radiation mechanism is purely Smith-Purcell one (possibly with an additional
Cherenkov radiation if the source moves in a medium). In the present paper
we consider a problem with two types of radiation mechanisms acting
together: synchrotron and Smith-Purcell ones. Namely, we will study the
radiation from a point charge rotating around a metallic diffraction grating
on a cylindrical surface, with the strips parallel to the cylinder axis.
Note that though the primary source of the radiation for both the
synchrotron and Smith-Purcell emissions is the electromagnetic field of the
charged particle, the Smith-Purcell radiation is formed by the medium as a
result of its dynamic polarization by the field of the moving charge. The
Smith-Purcell effect from an annular electron beam moving parallel to the
axis of a cylindrical grating, with the grooves perpendicular to the axis,
has been recently investigated in \cite{Blue15}. Of course, in this case the
only radiation mechanism is the Smith-Purcell one.

The outline of the paper is the following. In the next section we consider
the electromagnetic fields and the radiation from a charge rotating around a
solid conducting cylinder. The surface charge and current densities, induced
on the cylinder are determined. By using these densities, in section \ref%
{sec:Grat} we investigate the vector potential, the strengths of the
electric and magnetic fields for a charge rotating around a cylindrical
diffraction grating with metallic strips parallel to the cylinder axis. Then
we evaluate the spectral-angular density of the radiation intensity and
present the results of the numerical analysis for the number of the radiated
quanta. The main results of the paper are summarized in section \ref%
{sec:Conc}.

\section{Radiation from a charge rotating around a conducting cylinder}

\label{sec:Cyl}

In this section we consider the radiation of a point charge $q$, rotating
with the velocity $v$ along a circle of radius $r_{e}$ around an infinitely
long conducting cylinder with radius $r_{c}$, $r_{c}<r_{e}$ (see figure \ref%
{fig1}). For generality, we assume that the cylinder is immersed into a
homogeneous medium with the dielectric permittivity $\varepsilon $. The
corresponding electromagnetic field in the region outside the cylinder is
investigated in \cite{Kota00} (for the radiation from a charge rotating
inside a cylindrical waveguide with dielectric filling see \cite{Kota01}).
In accordance with the problem symmetry, we use cylindrical coordinates $%
(r,\varphi ,z)$ with the axis $z$ directed along the axis of the cylinder.
For the cylindrical coordinates of the charge we shall take $(r,\varphi
,z)=(r_{e},\omega _{0}t,0)$, where $\omega _{0}=v/r_{e}$ is the angular
velocity of the rotation.

\begin{figure}[tbph]
\begin{center}
\epsfig{figure=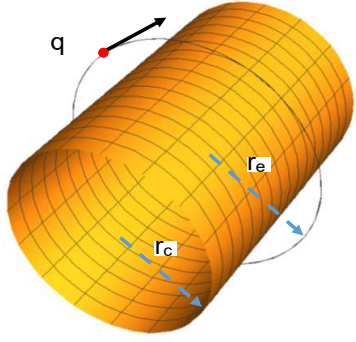,width=5.5cm,height=5.5cm}
\end{center}
\caption{Geometry of the problem corresponding to a point charge $q$
rotating around a conducting cylinder.}
\label{fig1}
\end{figure}

For the $l$th component of the vector potential one has the Fourier expansion%
\begin{equation}
A_{l}(x)=\sum\limits_{n=-\infty }^{+\infty }e^{in(\varphi -\omega
_{0}t)}\int_{-\infty }^{+\infty }dk_{z}\,e^{ik_{z}z}A_{nl}(k_{z}),
\label{Al}
\end{equation}%
where $x=(t,r,\varphi ,z)$ and $l=1,2,3$ correspond to the $r,\varphi ,z$
components, respectively. In the region $r>r_{c}$, these components are
decomposed as
\begin{equation}
A_{nl}(k_{z})=A_{nl}^{(0)}(k_{z})+A_{nl}^{(c)}(k_{z}),  \label{Adec}
\end{equation}%
where $A_{nl}^{(0)}(k_{z})$ corresponds to the field in the absence of the
cylinder and $A_{nl}^{(c)}(k_{z})$ is the field induced by the presence of
the cylinder. In the Lorentz gauge, for the separate contributions one has
\begin{eqnarray}
A_{nl}^{(0)}(k_{z}) &=&-\frac{vq}{4ci^{l-1}}\sum_{\alpha =\pm 1}\alpha
^{l}J_{n+\alpha }(\lambda r_{<})H_{n+\alpha }(\lambda r_{>}),  \notag \\
A_{nl}^{(c)}(k_{z}) &=&\frac{vq}{4ci^{l-1}}\sum_{\alpha =\pm 1}\alpha ^{l}%
\frac{H_{n+\alpha }(\lambda r_{0})}{H_{n+\alpha }(\lambda r_{1})}J_{n+\alpha
}(\lambda r_{1})H_{n+\alpha }(\lambda r),  \label{Anlc}
\end{eqnarray}%
for $l=1,2$, and $A_{n3}^{(0)}(k_{z})=A_{n3}^{(c)}(k_{z})=0$. Here, $J_{\nu
}(x)$ and $H_{\nu }(x)\equiv H_{\nu }^{(1)}(x)$ are the Bessel function and
the Hankel function of the first kind, $r_{<}=\mathrm{min}(r_{e},r)$, $r_{>}=%
\mathrm{max}(r_{e},r)$, and%
\begin{equation}
\lambda =\sqrt{n^{2}\omega _{0}^{2}\varepsilon /c^{2}-k_{z}^{2}}.
\label{lamb}
\end{equation}

Having the vector potential we can evaluate the electric and magnetic
fields. The corresponding Fourier expansions can be written in the form%
\begin{equation}
F_{l}(x)=\sum_{n=-\infty }^{\infty }e^{in(\varphi -\omega
_{0}t)}\int_{-\infty }^{+\infty }dk_{z}\,e^{ik_{z}z}F_{nl}(k_{z}),
\label{Fl}
\end{equation}%
where $F=E$ and $F=H$ for the electric and magnetic fields, respectively.
Considering the region $r>r_{e}$, for the Fourier components of the magnetic
field with $n\geqslant 0$ one gets%
\begin{eqnarray}
H_{nl}(k_{z}) &=&\frac{vqk_{z}}{4ci^{l-1}}\sum_{\alpha =\pm 1}\alpha
^{l-1}B_{n+\alpha }H_{n+\alpha }(\lambda r),\;l=1,2,  \notag \\
H_{n3}(k_{z}) &=&\frac{ivq\lambda }{4c}\sum_{\alpha =\pm 1}\alpha
B_{n+\alpha }H_{n}(\lambda r),  \label{Hnz}
\end{eqnarray}%
with the notation%
\begin{equation}
B_{n+\alpha }=J_{n+\alpha }(\lambda r_{e})-\frac{H_{n+\alpha }(\lambda r_{e})%
}{H_{n+\alpha }(\lambda r_{c})}J_{n+\alpha }(\lambda r_{c}).  \label{Bn}
\end{equation}%
For the components with $n<0$ we have the relation $H_{nl}(k_{z})=H_{-nl}^{%
\ast }(-k_{z})$. For $n\neq 0$, the electric field is found by using the
relation $\mathbf{E}_{n}=ic(\varepsilon n\omega _{0})^{-1}\mathrm{rot\,}%
\mathbf{H}_{n}$. This leads to the expressions%
\begin{eqnarray}
E_{nl}(k_{z}) &=&\frac{vq}{8i^{l}\varepsilon n\omega _{0}}\sum_{\alpha =\pm
1}\alpha ^{l}\left[ \left( n^{2}\omega _{0}^{2}\varepsilon
/c^{2}+k_{z}^{2}\right) B_{n+\alpha }-\lambda ^{2}B_{n-\alpha }\right]
H_{n+\alpha }(\lambda r),  \notag \\
E_{n3}(k_{z}) &=&\frac{vqk_{z}\lambda }{4\varepsilon n\omega _{0}}%
\sum_{\alpha =\pm 1}B_{n+\alpha }H_{n}(\lambda r),  \label{Enz}
\end{eqnarray}%
with $l=1,2$.

Similar to the vector potential, the Fourier components for the electric and
magnetic fields are presented as%
\begin{equation}
F_{nl}(k_{z})=F_{nl}^{(0)}(k_{z})+F_{nl}^{(c)}(k_{z}),  \label{Fdec}
\end{equation}%
where $F_{nl}^{(0)}(k_{z})$ are the corresponding components in the problem
without the cylinder and the part $F_{nl}^{(c)}(k_{z})$ is induced by the
conducting cylinder. The expressions for $F_{nl}^{(0)}(k_{z})$ are obtained
from the formulae (\ref{Hnz}) and (\ref{Enz}) by the replacement $%
B_{n+\alpha }\rightarrow J_{n+\alpha }(\lambda r_{e})$. The contribution in
the fields induced by the cylinder comes from the second term in the
right-hand side of (\ref{Bn}). The expressions for $F_{nl}^{(c)}(k_{z})$
have the form (\ref{Hnz}) and (\ref{Enz}) with the replacement $B_{n+\alpha
}\rightarrow -J_{n+\alpha }(\lambda r_{c})H_{n+\alpha }(\lambda
r_{e})/H_{n+\alpha }(\lambda r_{c})$. In the region $r_{c}\leqslant r<r_{e}$
the expressions for the parts $F_{nl}^{(c)}(k_{z})$ remain the same, whereas
in the expressions for $F_{nl}^{(0)}(k_{z})$ the replacements $%
J\rightleftarrows H$ should be made. Comparing the expressions (\ref{Anlc})
for the parts $A_{nl}^{(0)}(k_{z})$ and $A_{nl}^{(c)}(k_{z})$, we see that
for large values of $n$ the Fourier component of the field induced by the
cylinder in the region $r>r_{e}$ is approximated by the field of an
effective point charge
\begin{equation}
q_{\mathrm{im}}=-q\frac{r_{e}H_{n}(\lambda r_{e})}{r_{c}H_{n}(\lambda r_{c})}%
,  \label{qim}
\end{equation}%
located on the cylinder surface and having cylindrical coordinates $%
(r,\varphi ,z)=(r_{c},\omega _{0}t,0)$.

If the charge is close to the cylinder surface, $r_{e}/r_{c}-1\ll 1$, for $%
B_{n+\alpha }$, to the leading order, one has%
\begin{equation}
B_{n+\alpha }\approx -\frac{2i}{\pi \lambda r_{c}}\frac{r_{e}/r_{c}-1}{%
H_{n+\alpha }(\lambda r_{c})}.  \label{BnClose}
\end{equation}%
This shows that in the limit $r_{e}\rightarrow r_{c}$ the fields in the
region $r>r_{e}$ vanish. In this limit the field of the charge is
compensated by its image.

Having the electric field in the region $r_{c}\leqslant r<r_{e}$ we can find
the surface charge density $\sigma (\varphi ,z,t)$ on the cylinder. For the
corresponding Fourier transform, $\sigma _{n}(k_{z})$, defined in accordance
with%
\begin{equation}
\sigma (\varphi ,z,t)=\sum\limits_{n=-\infty }^{+\infty }e^{in(\varphi
-\omega _{0}t)}\int_{-\infty }^{+\infty }dk_{z}\,e^{ik_{z}z}\sigma
_{n}(k_{z}),  \label{sig}
\end{equation}%
one has $\sigma _{n}=\varepsilon E_{n1,r=r_{c}}(k_{z})/(4\pi )$. By using
the expression for the radial component of the electric field it can be seen
that
\begin{equation}
\sigma _{n}(k_{z})=-\frac{qr_{e}}{8\pi ^{2}r_{c}^{2}}\sum_{\alpha =\pm 1}%
\frac{H_{n+\alpha }(\lambda r_{e})}{H_{n+\alpha }(\lambda r_{c})}.
\label{sign}
\end{equation}%
Substituting into (\ref{sig}) and integrating over $\varphi $ and $z$, we
can show that%
\begin{equation}
r_{c}\int_{0}^{2\pi }d\varphi \int_{-\infty }^{+\infty }dz\,\sigma (\varphi
,z,t)=-q.  \label{TotCharge}
\end{equation}%
It is easy to see that, in the limit $r_{e}\rightarrow r_{c}$ for the
surface charge (\ref{sig}) one gets $\sigma (\varphi
,z,t)|_{r_{e}\rightarrow r_{c}}=-q\delta \left( \varphi -\omega _{0}t\right)
\delta (z)/r_{c}$, where we have assumed that $0\leqslant \varphi -\omega
_{0}t<2\pi $. The latter corresponds to a point charge $-q$ located on the
cylinder surface.

For the surface current density induced on the cylinder surface one has $%
\mathbf{j}_{s}=c[\mathbf{n}\times \mathbf{H}]/(4\pi )|_{r=r_{c}}$, where $%
\mathbf{n}$ is the normal to the cylinder. By using the expressions (\ref%
{Hnz}), we can see that the only nonzero component of the current density is
along the azimuthal direction, $j_{sl}=0$ for $l=1,3$, and $%
j_{s2}=-cH_{3,r=r_{c}}/(4\pi )$. From (\ref{Hnz}), for the corresponding
Fourier component of the magnetic field one gets%
\begin{equation}
H_{n3}(k_{z})|_{r=r_{c}}=\frac{vq}{2\pi cr_{c}}\sum_{\alpha =\pm 1}\frac{%
H_{n+\alpha }(\lambda r_{e})}{H_{n+\alpha }(\lambda r_{c})}.  \label{Hn3}
\end{equation}%
This shows that the Fourier components of the surface charge and current
densities are related by the standard formula $j_{sn2}(k_{z})=v^{\prime
}\sigma _{n}(k_{z})$, where $v^{\prime }=\omega _{0}r_{c}=vr_{c}/r_{e}$ is
the velocity of the charge image on the cylinder surface.

At large distances from the cylinder, the angular density of the radiation
intensity at the frequency $n\omega _{0}$, $n=1,2,\ldots $, is given by the
expression%
\begin{equation}
\frac{dI_{n}^{(c)}}{d\Omega }=\frac{q^{2}n^{2}\omega _{0}^{2}}{8\pi \sqrt{%
\varepsilon }c}\beta ^{2}\left[ \left\vert B_{n+1}-B_{n-1}\right\vert
^{2}+\left\vert B_{n+1}+B_{n-1}\right\vert ^{2}\cos ^{2}\theta \right] ,
\label{dIn}
\end{equation}%
where $\beta =v\sqrt{\varepsilon }/c$, $d\Omega =\sin \theta d\theta
d\varphi $ is the solid angle element, $\theta $ is the angle between the
radiation direction and the axis $z$, and
\begin{equation}
k_{z}=\frac{n\omega _{0}}{c}\sqrt{\varepsilon }\cos \theta ,  \label{kz}
\end{equation}%
is the projection of the wave vector on the cylinder axis. For $B_{n\pm 1}$
in (\ref{dIn}) one has the expression (\ref{Bn}) with%
\begin{equation}
\lambda =\frac{n\omega _{0}}{c}\sqrt{\varepsilon }\sin \theta ,
\label{lamb1}
\end{equation}%
and, hence, $\lambda r_{e}=n\beta \sin \theta $. Let us consider some
limiting cases of the formula (\ref{dIn}).

In the absence of the cylinder one has $B_{n+\alpha }=J_{n+\alpha }(\lambda
r_{e})$ and (\ref{dIn}) is reduced to the corresponding expression for the
synchrotron radiation in a homogeneous and isotropic medium (see \cite%
{Tsyt51}):%
\begin{equation}
\frac{dI_{n}^{(0)}}{d\Omega }=\frac{q^{2}n^{2}\omega _{0}^{2}}{2\pi c\sqrt{%
\varepsilon }}\left[ \beta ^{2}J_{n}^{\prime 2}(n\beta \sin \theta )+\cot
^{2}\theta J_{n}^{2}(n\beta \sin \theta )\right] .  \label{dIn0}
\end{equation}%
By using the asymptotic expression (\ref{BnClose}), in the limit $%
r_{e}\rightarrow r_{c}$ we see that the radiation intensity from a charge
rotating around a conducting cylinder vanishes as $\left(
r_{e}/r_{c}-1\right) ^{2}$. In this limit the field of the charge is
compensated by the field of its image on the cylinder surface. For small
angles $\theta $, to the leading order one gets%
\begin{equation}
\frac{dI_{n}^{(c)}}{d\Omega }\approx \left[ 1-\left( r_{c}/r_{e}\right)
^{2\left( n-1\right) }\right] ^{2}\frac{dI_{n}^{(0)}}{d\Omega },
\label{SmallAng}
\end{equation}%
with%
\begin{equation}
\frac{dI_{n}^{(0)}}{d\Omega }\approx \frac{q^{2}\omega _{0}^{2}}{\pi c\sqrt{%
\varepsilon }}\frac{(n\beta /2)^{2n}}{\Gamma ^{2}(n)}\sin ^{2(n-1)}\theta .
\label{SmallAng0}
\end{equation}%
In a homogeneous medium, the only nonzero contribution to the radiation at
the angle $\theta =0$ comes from the harmonic $n=1$. In the problem with the
conducting cylinder, the radiation intensity at zero angle vanishes for all
harmonics including $n=1$.

For large values of the radiation harmonic $n$ and for $x<1$, we can use the
asymptotic expressions%
\begin{eqnarray}
J_{n+\alpha }(nx) &\approx &\frac{\eta _{1}^{1/2}(x)x^{-|\alpha |}}{\pi
\left( 1-x^{2}\right) ^{1/4}}\left[ K_{1/3}(n\eta _{1}(x))-\alpha \sqrt{%
1-x^{2}}K_{2/3}(n\eta _{1}(x))\right] ,  \notag \\
Y_{n+\alpha }(nx) &\approx &-\frac{\eta _{1}^{1/2}(x)x^{-|\alpha |}}{\left(
1-x^{2}\right) ^{1/4}}\left[ I_{1/3}^{(+)}(n\eta _{1}(x))+\alpha \sqrt{%
1-x^{2}}I_{2/3}^{(+)}(n\eta _{1}(x))\right] ,  \label{JYas}
\end{eqnarray}%
where $\alpha =0,\pm 1$, $Y_{\nu }(u)$ is the Neumann function, $I_{\nu
}^{(+)}(u)=I_{\nu }(u)+I_{-\nu }(u)$,
\begin{equation}
\eta _{1}(x)=\ln \frac{1+\sqrt{1-x^{2}}}{x}-\sqrt{1-x^{2}},  \label{eta}
\end{equation}%
with $K_{\nu }(u)$ and $I_{\nu }(u)$ being the modified Bessel functions.
For $x>1$ the asymptotic formulas are given by
\begin{eqnarray}
J_{n+\alpha }(nx) &\approx &\frac{\eta _{2}^{1/2}(x)x^{-|\alpha |}}{\sqrt{3}%
\left( x^{2}-1\right) ^{1/4}}\left[ J_{1/3}^{(+)}(n\eta _{2}(x))+\alpha
\sqrt{x^{2}-1}J_{2/3}^{(-)}(n\eta _{2}(x))\right] ,  \notag \\
Y_{n+\alpha }(nx) &\approx &\frac{\eta _{2}^{1/2}(x)x^{-|\alpha |}}{\left(
x^{2}-1\right) ^{1/4}}\left[ J_{1/3}^{(-)}(n\eta _{2}(x))-\alpha \sqrt{%
x^{2}-1}J_{2/3}^{(+)}(n\eta _{2}(x))\right] ,  \label{JYas2}
\end{eqnarray}%
with $\alpha =0,\pm 1$, $J_{\nu }^{(\pm )}(u)=J_{\nu }(u)\pm J_{-\nu }(u)$,
and%
\begin{equation}
\eta _{2}(x)=\sqrt{x^{2}-1}-\arccos \left( 1/x\right) .  \label{eta1}
\end{equation}%
The asymptotic formulas (\ref{JYas}) and (\ref{JYas2}) are obtained from the
uniform asymptotic expansions for the cylinder functions, given, for
example, in \cite{Abra72}, by using the expressions for the Airy functions
and their derivatives in terms of the cylinder functions with the orders $%
\pm 1/3$ and $\pm 2/3$. Note that $\eta _{j}(1)=0$ for $j=1,2$. The function
$\eta _{1}(x)$ ($\eta _{2}(x)$) is a monotonically decreasing (increasing)
function in the region $0<x\leqslant 1$ ($1\leqslant x<\infty $). For $%
|x-1|\ll 1$ one has
\begin{equation}
\eta _{j}(x)\approx \frac{2^{3/2}}{3}|x-1|^{3/2}.  \label{etaas}
\end{equation}%
The asymptotic expressions for the Hankel function are obtained from (\ref%
{JYas}) and (\ref{JYas2}), by taking into account that $H_{\nu }(x)=J_{\nu
}(x)+iY_{\nu }(x)$.

For $\beta \sin \theta <1$, the radiation from a charge rotating in a
homogeneous medium is emitted on the harmonics $n\lesssim 1/\eta _{1}(\beta
\sin \theta )$. For $n\gg 1/\eta _{1}(\beta \sin \theta )$ the corresponding
radiation intensity is suppressed by the factor $e^{-2n\eta _{1}(\beta \sin
\theta )}$. Under the condition $\beta >1$ and for $\beta \sin \theta >1$,
by using the asymptotic expressions (\ref{JYas2}), one gets
\begin{equation}
\frac{dI_{n}^{(0)}}{d\Omega }\approx \frac{q^{2}n^{2}\omega _{0}^{2}}{6\pi c%
\sqrt{\varepsilon }}\frac{\eta _{2}(y)}{\sin ^{2}\theta }\left[ \left(
y^{2}-1\right) ^{1/2}J_{2/3}^{(-)2}(n\eta _{2}(y))+\cos ^{2}\theta \frac{%
J_{1/3}^{(+)2}(n\eta _{2}(y))}{\left( y^{2}-1\right) ^{1/2}}\right] ,
\label{In0as}
\end{equation}%
with $y=\beta \sin \theta $. In this case the radiation intensity increases
with increasing $n$. However, for large values of $n$ the dispersion of the
dielectric permittivity should be taken into account, $\varepsilon
=\varepsilon (n\omega _{0})$, and the increase of the intensity is
restricted by the Cherenkov condition $v\varepsilon (n\omega _{0})/c>1$.

For a charge rotating around a conducting cylinder and under the condition $%
\beta \sin \theta <1$, the cylinder-induced effects in the radiation
intensity on the harmonics $n\gg 1/\eta _{1}(\beta \sin \theta )$, compared
with $dI_{n}^{(0)}/d\Omega $, are suppressed by an additional factor $%
e^{-2n[\eta _{1}(\beta ^{\prime }\sin \theta )-\eta _{1}(\beta \sin \theta
)]}$. Note that for the ratio $r_{c}/r_{e}$ close to 1 one has%
\begin{equation}
\eta _{1}(\beta ^{\prime }\sin \theta )-\eta _{1}(\beta \sin \theta )\approx
\sqrt{1-\beta ^{2}\sin ^{2}\theta }(1-r_{c}/r_{e}).  \label{etaclose}
\end{equation}%
In this case the relative suppression factor can be presented as%
\begin{equation}
e^{-2n[\eta _{1}(\beta ^{\prime }\sin \theta )-\eta _{1}(\beta \sin \theta
)]}\approx \exp [-4\pi \sqrt{\beta ^{-2}-\sin ^{2}\theta }%
(r_{e}-r_{c})/\lambda _{r}],  \label{supfactor}
\end{equation}%
where $\lambda _{r}$ is the radiation wavelength. For the radiation in the
vacuum ($\varepsilon =1$) and for the radiation angles $\theta $ not too
close to $\pi /2$, the effects induced by the cylinder are exponentially
suppressed for wavelengths $\lambda _{r}<r_{e}-r_{c}$. For relativistic
particles and for angles close to $\pi /2$ one has $\beta ^{-2}-\sin
^{2}\theta \approx \gamma ^{-2}+\left( \theta -\pi /2\right) ^{2}$ and the
contribution of the cylinder to the total radiation intensity can be
essential for wavelengths much smaller than the distance from the cylinder.

\begin{figure}[tbph]
\begin{center}
\begin{tabular}{cc}
\epsfig{figure=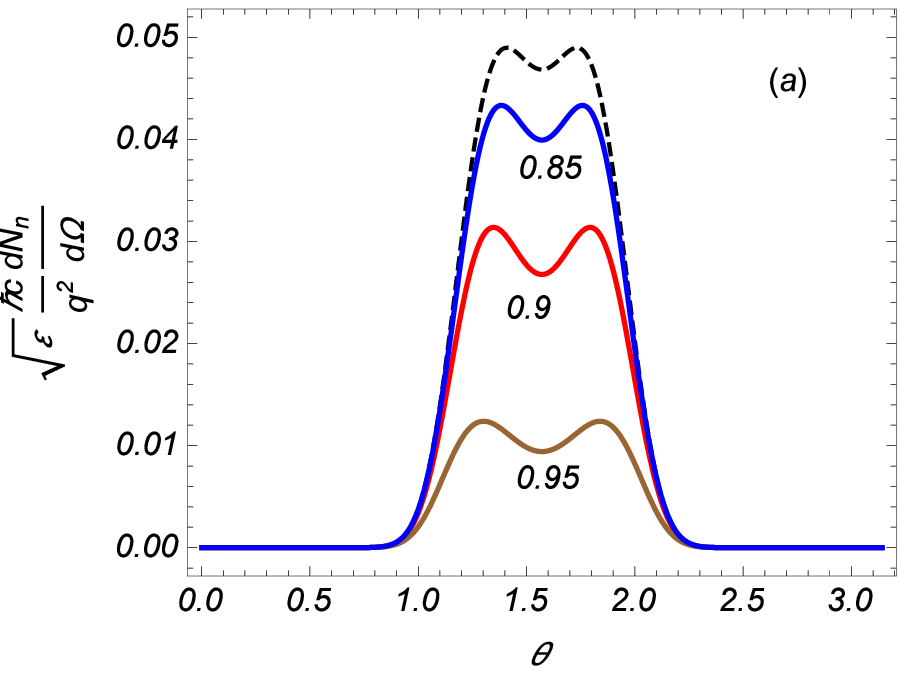,width=7.5cm,height=6.cm} & \quad %
\epsfig{figure=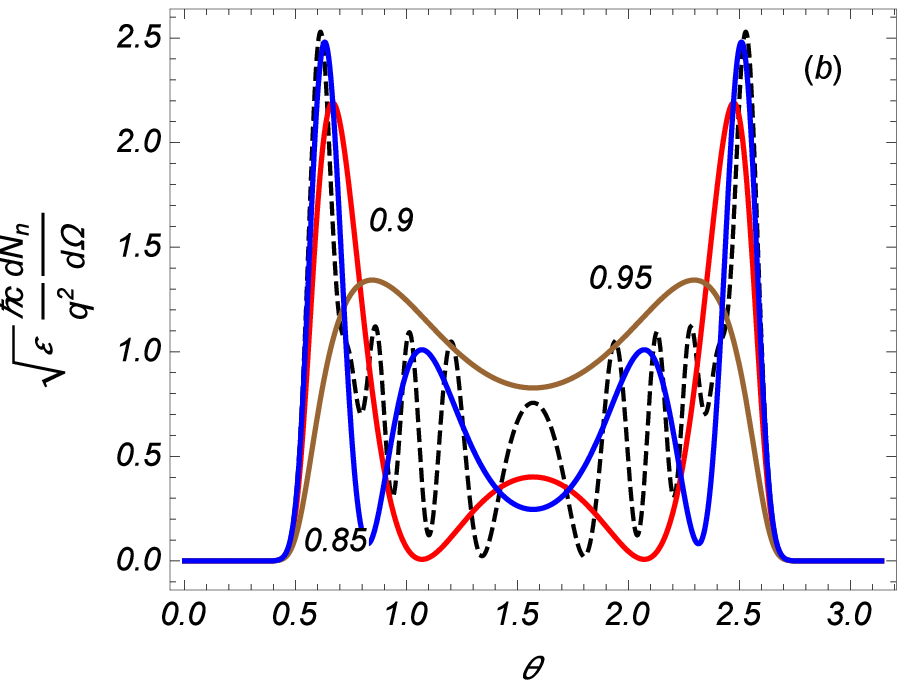,width=7.5cm,height=6.cm}%
\end{tabular}%
\end{center}
\caption{Angular density of the number of the radiated quanta per period of
the charge rotation versus the radiation angle with respect to the cylinder
axis. The graphs are plotted for the electron energy 2 MeV and for the
radiation harmonic $n=25$. The numbers near the curves are the values of $%
r_{c}/r_{e}$ and the dashed curves correspond to the radiation in the
absence of the cylinder. For the panels (a) and (b) we have taken $\protect%
\varepsilon =1$ and $\protect\varepsilon =3.75$, respectively.}
\label{fig2}
\end{figure}

In the figures below we plot the angular density for the number of the
quanta on a given harmonic radiated per period of the charge rotation:%
\begin{equation}
\frac{dN_{n}^{(c)}}{d\Omega }=\frac{T}{\hbar n\omega _{0}}\frac{dI_{n}^{(c)}%
}{d\Omega }.  \label{dN}
\end{equation}%
Figure \ref{fig2} presents this quantity as a function of the angle $\theta $
(in radians) for the harmonic $n=25$ and for the electron energy $E_{e}=2$
MeV. The numbers near the curves are the values of the ratio $r_{c}/r_{e}$.
The dashed curves correspond to the radiation in the absence of a conducting
cylinder. The panels (a) and (b) present the cases $\varepsilon =1$
(radiation in the vacuum) and $\varepsilon =3.75$ (dielectric permittivity
for quartz). As is seen, in the second case the number of the radiated
quanta is essentially larger compared with the case of the radiation in the
vacuum. This is related to the fact that, in addition to the synchrotron
radiation, one has also Cherenkov radiation. For the case of the panel (b),
the radiation is mainly located in the angular region $\pi /2-\theta
_{C}<\theta <\pi /2+\theta _{C}$, where $\theta _{C}=\arccos (1/\beta )$.
For the parameters corresponding to this panel one has $\pi /2-\theta
_{C}\approx 0.56$, whereas for the left peak one has $\theta \approx 0.61$.
The number of oscillations in the angular range $\pi /2-\theta _{C}<\theta
<\pi /2+\theta _{C}$ increases with increasing $n$. For the case of the
radiation in the vacuum, the number of the radiated quanta is decreasing
with increasing $r_{c}/r_{e}$. As it has been already explained above, the
radiation intensity vanishes in the limit $r_{c}/r_{e}\rightarrow 1$.

\begin{figure}[tbph]
\begin{center}
\epsfig{figure=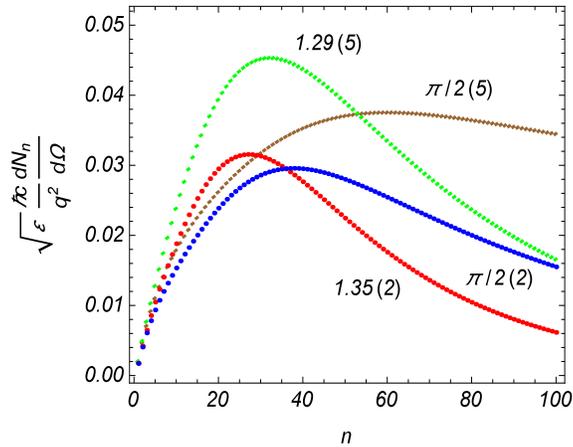,width=7.5cm,height=6.cm}
\end{center}
\caption{Angular density of the number of the radiated quanta as a function
of the radiation harmonic for the energies of the electron 2 MeV and 5 MeV
(numbers in the brackets) and for $r_{c}/r_{e}=0.9$, $\protect\varepsilon =1$%
. The numbers near the curves correspond to the values of $\protect\theta $.}
\label{fig3}
\end{figure}

Figure \ref{fig3} presents the number of the radiated quanta as a function
of the harmonic $n$ for the case of the radiation in the vacuum ($%
\varepsilon =1$). The graphs are plotted for $r_{c}/r_{e}=0.9$. The numbers
near the curves are the corresponding values of the angle $\theta $ and the
numbers in the brackets correspond to the value of the energy in MeVs. We
have considered the radiation in the plane of the radiation ($\theta =\pi /2$%
) and along the angles at which the angular density for the number of the
radiated quanta has a maximum for the harmonic $n=25$.

\section{Electromagnetic fields and radiation intensity in the geometry of a
cylindrical grating}

\label{sec:Grat}

\begin{figure}[tbph]
\begin{center}
\epsfig{figure=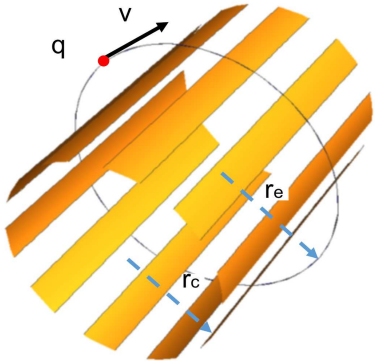,width=6.cm,height=6.cm}
\end{center}
\caption{Point charge rotating around a cylindrical grating.}
\label{fig4}
\end{figure}

In this section we consider the radiation from a charge rotating around a
diffraction grating on the cylindrical surface with radius $r_{c}$. The
grating consists infinitely long metallic strips with width $a$ and with the
separation $b$ (see figure \ref{fig4}). The problem is mathematically
complicated and an exact analytical result is not available. This is already
the case for simpler problems of the Smith-Purcell radiation from planar
gratings. For the latter geometry various approximate analytic and numerical
methods have been developed for the evaluation of the spectral-angular
distribution of the radiation intensity (for the comparison of different
models in the calculations of the Smith-Purcell radiation intensity see \cite%
{Karl06}). On the base of the analysis given above, we approximate the
effect of the grating on the radiation intensity by the surface currents
induced on the strips by the field of the rotating charge. For planar
gratings this method has been discussed in \cite{Wals94,Brow98} (for further
developments of the method see \cite{Karl09}).

For the current density induced on the strips of the cylindrical grating one
has%
\begin{equation}
j_{l}^{(\mathrm{s})}=v_{l}^{\prime }\sigma ^{(\mathrm{s})}(\varphi
,z,t)\delta (r-r_{c}),  \label{jlim}
\end{equation}%
where $v_{l}^{\prime }=v^{\prime }\delta _{2l}$. For the surface charge
density we use the expression%
\begin{equation}
\sigma ^{(\mathrm{s})}(\varphi ,z,t)=\left\{
\begin{array}{cc}
\sigma (\varphi ,z,t), & m\varphi _{1}\leq \varphi \leq m\varphi
_{1}+\varphi _{0} \\
0, & \text{otherwise}%
\end{array}%
\right. ,  \label{SurfCharge}
\end{equation}%
where $m=0,1,2,...,N-1$, with
\begin{equation}
N=\frac{2\pi r_{c}}{a+b},  \label{Ngr}
\end{equation}%
being the number of the periods of the grating. In (\ref{SurfCharge}), the
function $\sigma (\varphi ,z,t)$ is given by (\ref{sig}) with $\sigma
_{n}(k_{z})$ from (\ref{sign}) and
\begin{equation}
\varphi _{0}=a/r_{c},\;\varphi _{1}=\left( a+b\right) /r_{c},  \label{phi0}
\end{equation}%
are the angular width of the strips and the angular period of the grating,
respectively.

\subsection{Electric and magnetic fields}

The vector potential for the field generated by the surface current density (%
\ref{jlim}) is determined from the formula%
\begin{equation}
A_{i}^{(\mathrm{s})}(\mathbf{r},t)=-\frac{1}{2\pi ^{2}c}\int G_{il}(\mathbf{r%
},t,\mathbf{r}^{\prime },t^{\prime })j_{l}^{(\mathrm{s})}(\mathbf{r}^{\prime
},t^{\prime })d\mathbf{r}^{\prime }dt^{\prime },  \label{Aim}
\end{equation}%
where $G_{il}(\mathbf{r},t,\mathbf{r}^{\prime },t^{\prime })$ is the
electromagnetic field Green function in a homogeneous medium with
permittivity $\varepsilon $ and summation over $l$ is understood. In
accordance with the problem symmetry, for the Green function we have the
following Fourier expansion:%
\begin{equation}
G_{il}(\mathbf{r},t,\mathbf{r}^{\prime },t^{\prime })=\sum_{n=-\infty
}^{+\infty }\int_{-\infty }^{+\infty }dk_{z}\int_{-\infty }^{+\infty
}d\omega \,G_{il}(n,k_{z},\omega ,r,r^{\prime })e^{in(\varphi -\varphi
^{\prime })+ik_{z}(z-z^{\prime })-i\omega (t-t^{\prime })}.  \label{GilF}
\end{equation}%
Substituting in (\ref{Aim}) and by taking into account (\ref{jlim}), after
redefining $n\rightarrow n+Nm$, one gets%
\begin{eqnarray}
A_{l}^{(\mathrm{s})}(\mathbf{r},t) &=&-\frac{4\pi v^{\prime }r_{c}}{c}%
\sum_{n,m=-\infty }^{+\infty }e^{-iNm\varphi _{0}/2}S_{m}e^{i\left(
n-Nm\right) \varphi -in\omega _{0}t}  \notag \\
&&\times \int_{-\infty }^{+\infty
}dk_{z}\,e^{ik_{z}z}G_{l2}(n+Nm,k_{z},n\omega _{0},r,r_{c})\sigma
_{n}(k_{z}),  \label{Ais}
\end{eqnarray}%
with the notation%
\begin{equation}
S_{m}=\frac{1}{\pi m}\sin \left( \frac{\pi ma}{a+b}\right) .  \label{Sms}
\end{equation}

For the Fourier components of the Green tensor, appearing in (\ref{Ais}),
one has
\begin{equation}
G_{l2}(n+Nm,k_{z},n\omega _{0},r,r_{c})=\frac{\pi }{4}i^{1-l}\sum_{\alpha
=\pm 1}\alpha ^{l}J_{n+Nm+\alpha }(\lambda r_{<}^{\prime })H_{n+Nm+\alpha
}(\lambda r_{>}^{\prime }),  \label{Gl2}
\end{equation}%
for $l=1,2$ and $G_{32}(n,k_{z},\omega ,r,r_{c})=0$. Here, $\lambda $ is
defined by (\ref{lamb}) and $r_{<}^{\prime }=\mathrm{min}(r_{c},r)$, $%
r_{>}^{\prime }=\mathrm{max}(r_{c},r)$. Substituting (\ref{Gl2}) into (\ref%
{Ais}), the components of the vector potential are presented as the Fourier
expansion%
\begin{equation}
A_{l}^{(\mathrm{s})}(x)=\sum_{n,m=-\infty }^{+\infty }\int_{-\infty
}^{+\infty }dk_{z}\,e^{i\left( n+mN\right) \varphi -in\omega
_{0}t+ik_{z}z}A_{nml}^{(\mathrm{s})}(k_{z},r),  \label{Ais1}
\end{equation}%
where%
\begin{equation}
A_{nml}^{(\mathrm{s})}(k_{z},r)=\frac{i^{1-l}qv}{4c}e^{-iNm\varphi
_{0}/2}S_{m}\sum_{\alpha =\pm 1}\alpha ^{l}\frac{H_{n+\alpha }(\lambda r_{e})%
}{H_{n+\alpha }(\lambda r_{c})}J_{n+mN+\alpha }(\lambda r_{<}^{\prime
})H_{n+mN+\alpha }(\lambda r_{>}^{\prime }),  \label{Ais2}
\end{equation}%
for $l=1,2$ and $A_{nm3}^{(\mathrm{s})}(k_{z},r)=0$. Compared to the case of
a solid cylinder, the Fourier expansion for the geometry of a grating
contains an additional summation over $m$. The latter corresponds to the
periodic structure along the azimuthal direction. As a result, the vector
potential for the total field is presented as
\begin{equation}
A_{l}(x)=A_{l}^{(0)}(x)+A_{l}^{(\mathrm{s})}(x),  \label{AnlSum}
\end{equation}%
where $A_{l}^{(0)}(x)$ is the corresponding field in a homogeneous medium
with permittivity $\varepsilon $.

The expansions similar to (\ref{Ais1}) can be written for the field
strengths:%
\begin{equation}
F_{l}^{(\mathrm{s})}(x)=\sum_{n,m=-\infty }^{+\infty }\int_{-\infty
}^{+\infty }dk_{z}\,e^{i\left( n+mN\right) \varphi -in\omega
_{0}t+ik_{z}z}F_{nml}^{(\mathrm{s})}(k_{z},r),  \label{Fsl}
\end{equation}%
with $F=E$ and $F=H$ for the electric and magnetic fields, respectively. For
the Fourier components of the magnetic field in the region $r>r_{c}$ one gets%
\begin{eqnarray}
H_{nml}^{(\mathrm{s})}(k_{z},r) &=&\frac{qvk_{z}}{4ci^{l-1}}\sum_{\alpha
=\pm 1}\alpha ^{l-1}B_{n,m}^{(\alpha )}H_{n+mN+\alpha }(\lambda r),  \notag
\\
H_{nm3}^{(\mathrm{s})}(k_{z},r) &=&\frac{iqv\lambda }{4c}\sum_{\alpha =\pm
1}\alpha B_{n,m}^{(\alpha )}H_{n+mN}(\lambda r).  \label{Hs}
\end{eqnarray}%
with $l=1,2$ and with the notation%
\begin{equation}
B_{n,m}^{(\alpha )}=-e^{-\frac{\pi ia}{a+b}m}S_{m}\frac{H_{n+\alpha
}(\lambda r_{e})}{H_{n+\alpha }(\lambda r_{c})}J_{n+mN+\alpha }(\lambda
r_{c}).  \label{Bnm}
\end{equation}%
For $n\neq 0$, the spectral components of the electric field are found from $%
\mathbf{E}_{n}=ic(\varepsilon n\omega _{0})^{-1}\mathrm{rot\,}\mathbf{H}_{n}$%
. For the Fourier components, this gives the expressions%
\begin{eqnarray}
E_{nml}^{(\mathrm{s})}(k_{z},r) &=&\frac{i^{-l}qv}{8\varepsilon n\omega _{0}}%
\sum_{\alpha =\pm 1}\alpha ^{l}\left[ \left( n^{2}\omega _{0}^{2}\varepsilon
/c^{2}+k_{z}^{2}\right) B_{n,m}^{(\alpha )}-\lambda ^{2}B_{n,m}^{(-\alpha )}%
\right] H_{n+mN+\alpha }(\lambda r),  \notag \\
E_{nm3}^{(\mathrm{s})}(k_{z},r) &=&\frac{qvk_{z}\lambda }{4\varepsilon
n\omega _{0}}\sum_{\alpha =\pm 1}B_{n,m}^{(\alpha )}H_{n+mN}(\lambda r),
\label{Es}
\end{eqnarray}%
where $l=1,2$.

We can write the expansion of the type (\ref{Fsl}) for the total fields%
\begin{equation}
F_{l}(x)=F_{l}^{(0)}(x)+F_{l}^{(\mathrm{s})}(x).  \label{Flt}
\end{equation}%
In the region $r>r_{e}$, the expressions for the corresponding Fourier
components are obtained from (\ref{Hs}) and (\ref{Es}) by the replacement $%
B_{n,m}^{(\alpha )}\rightarrow D_{n,m}^{(\alpha )}$ with%
\begin{equation}
D_{n,m}^{(\alpha )}=\delta _{0m}J_{n+\alpha }(\lambda
r_{e})+B_{n,m}^{(\alpha )}.  \label{Dnm}
\end{equation}%
Here the first term in the right-hand side comes from the part of the field
corresponding to the problem with a homogeneous medium when the cylindrical
grating is absent. In the region $r_{c}<r\,<r_{e}$, the expressions for $%
F_{l}^{(\mathrm{s})}(x)$ remain the same, whereas in the expressions for the
part $F_{l}^{(0)}(x)$ the replacement $J\rightleftarrows H$ of the Bessel
and Hankel functions should be made.

\subsection{Radiation intensity}

Having the fields, we can evaluate the radiation intensity from a charge
rotating around a diffraction grating. The average energy flux per unit time
through the cylindrical surface of radius $r$, coaxial with the axis of the
grating, is given by
\begin{equation}
I^{(g)}=\frac{c}{4\pi T}\int_{0}^{2\pi }d\varphi \int_{0}^{T}dt\int_{-\infty
}^{\infty }dz\,r\mathbf{n}_{r}\cdot \left[ \mathbf{E}\times \mathbf{H}\right]
,  \label{I1}
\end{equation}%
where $T=2\pi /\omega _{0}$ and $\mathbf{n}_{r}$ is the unit vector along
the radial direction $r$. Substituting the Fourier expansions of the type (%
\ref{Fsl}) for the electric and magnetic fields, we obtain the following
representation%
\begin{equation}
I^{(g)}=\pi cr\sum\limits_{n,m=-\infty }^{+\infty }\int_{-\infty }^{+\infty
}dk_{z}\,\left[ E_{nm2}(k_{z},r)H_{nm3}^{\ast
}(k_{z},r)-E_{nm3}(k_{z},r)H_{nm2}^{\ast }(k_{z},r)\right] ,  \label{Ig}
\end{equation}%
where the Fourier components are given by the expressions obtained from (\ref%
{Hs}) and (\ref{Es}) with the replacement $B_{n,m}^{(\alpha )}\rightarrow
D_{n,m}^{(\alpha )}$.

For large values of $r$, by using the asymptotic expressions for the Hankel
function for large arguments, we see that for the radiation field $\lambda
^{2}>0$ that corresponds to the integration region $k_{z}^{2}<n^{2}\omega
_{0}^{2}\sqrt{\varepsilon }/c^{2}$. In this region, introducing a new
integration variable $\theta $ in accordance with (\ref{kz}), the radiation
intensity is presented in the form%
\begin{equation}
I^{(g)}=2\pi \sum\limits_{n=1}^{\infty }\int_{0}^{\pi }d\theta \,\sin \theta
\frac{dI_{n}^{(g)}}{d\Omega },  \label{Ig1}
\end{equation}%
with the angular density of the radiation intensity at a given harmonic $%
n\omega _{0}$:%
\begin{equation}
\frac{dI_{n}^{(g)}}{d\Omega }=\frac{q^{2}\beta ^{2}n^{2}\omega _{0}^{2}}{%
8\pi c\sqrt{\varepsilon }}\sum_{m=-\infty }^{+\infty }\left[ \left\vert
R_{n,m}^{(+1)}-R_{n,m}^{(-1)}\right\vert ^{2}+\cos ^{2}\theta \left\vert
R_{n,m}^{(+1)}+R_{n,m}^{(-1)}\right\vert ^{2}\right] .  \label{dIg}
\end{equation}%
Here, as before, $\theta $ is the angle between the radiation direction and
the axis $z$, and
\begin{equation}
R_{n,m}^{(\alpha )}=\delta _{0m}J_{n+\alpha }(n\beta \sin \theta )-S_{m}%
\frac{H_{n+\alpha }(n\beta \sin \theta )}{H_{n+\alpha }(n\beta ^{\prime
}\sin \theta )}J_{n+mN+\alpha }(n\beta ^{\prime }\sin \theta ),  \label{Dnm1}
\end{equation}%
with $\beta ^{\prime }=r_{c}\beta /r_{e}$. Note that $R_{n,m}^{(\alpha
)}=e^{-\frac{i\pi m}{1+b/a}}D_{n,m}^{(\alpha )}$. The part in the radiation
intensity coming from the first term in the right-hand side of (\ref{Dnm1})
corresponds to the radiation in a homogeneous medium. The remaining parts
are induced by the diffraction grating.

For $a=0$ one has $S_{m}=0$ and (\ref{dIg}) is reduced to $%
dI_{n}^{(0)}/d\Omega $, given by (\ref{dIn0}). Another special case $b=0$
corresponds to a conducting cylinder. In this case, by taking into account
that $S_{m}=\delta _{0m}$, we see that $D_{n,m}^{(\alpha )}=\delta
_{0m}B_{n}^{(\alpha )}$ and the radiation intensity (\ref{dIg}) coincides
with (\ref{dIn}). If the charge trajectory is close to the grating, to the
leading order, in (\ref{Dnm1}) we can put $\beta ^{\prime }=\beta $ and the
radiation intensity is simplified to
\begin{eqnarray}
&&\left. \frac{dI_{n}^{(g)}}{d\Omega }\right\vert _{r_{1}=r_{0}}=\left(
\frac{b}{a+b}\right) ^{2}\frac{dI_{n}^{(0)}}{d\Omega }+\frac{%
q^{2}n^{2}\omega _{0}^{2}}{2\pi c\sqrt{\varepsilon }}\sum_{m\neq 0}S_{m}^{2}
\notag \\
&&\qquad \times \left[ \beta ^{2}J_{n+mN}^{\prime 2}(n\beta \sin \theta
)+\cot ^{2}\theta \left( 1+mN/n\right) ^{2}J_{n+mN}^{2}(n\beta \sin \theta )%
\right] .  \label{dIgclose}
\end{eqnarray}%
As it has been mentioned above, under the condition $\beta \sin \theta <1$,
for a given direction $\theta $ of the radiation and for the harmonics $n\gg
\eta _{1}(\beta \sin \theta )$, the first term in the right-hand side of (%
\ref{dIgclose}) is suppressed by the factor $e^{-2n\eta _{1}(\beta \sin
\theta )}$. This may not be the case for the second term corresponding to
the contribution of the Smith-Purcell radiation. Indeed, mathematically, the
smallness of the part $dI_{n}^{(0)}/d\Omega $ is related to the fact that
under the specified conditions the Bessel function $J_{n}(n\beta \sin \theta
)$ is exponentially small. The Smith-Purcell part of the radiation contains
the Bessel functions with the orders $n+mN$. Under the condition $|n+mN|\ll
n\beta \sin \theta $ one has $J_{n+mN}(n\beta \sin \theta )\sim 1/\sqrt{n}$
and the corresponding contributions to (\ref{dIgclose}) are not small.

As it has been mentioned in the previous section, for a charge rotating in
homogeneous medium or for a charge rotating around a conducting cylinder,
the angular density of the radiation intensity for the harmonics with $n>1$
vanishes in the limit $\theta \rightarrow 0$. This may not be the case in
the presence of the diffraction grating. For example, if $N>2$ and $(n\pm
1)/N$ is an integer (note that only one of this numbers can be an integer),
then for small angles the dominant contribution to (\ref{dIg}) comes from
the term with $m=-(n\pm 1)/N$ and to the leading order one has%
\begin{equation}
\frac{dI_{n}^{(g)}}{d\Omega }\approx \frac{q^{2}\beta ^{2}n^{2}\omega
_{0}^{2}}{4\pi \sqrt{\varepsilon }c}S_{(n\pm 1)/N}^{2}(r_{1}/r_{0})^{2n\pm
2}.  \label{dIgsmtet}
\end{equation}%
The corresponding radiation intensity in a homogeneous medium vanishes as $%
\sin ^{2n-2}(\theta )$ (see (\ref{SmallAng0})).

In the geometry of diffraction grating the behavior of the radiation
intensity on large values of the harmonic $n$ can be essentially different
from that for a charge rotating in the vacuum or around a solid cylinder. As
an illustration let us consider the charge rotating in the vacuum ($%
\varepsilon =1$) around the grating. In this case one has $\beta \sin \theta
<1$ and for large values of $n$ for the cylinder functions in (\ref{Dnm1})
with the orders $n+\alpha $ we can use the asymptotic formulas (\ref{JYas}).
In particular, for the harmonics $n\gg 1/\eta _{1}(\beta \sin \theta )$ the
part corresponding to the pure synchrotron radiation (coming from the first
term in the right-hand side of (\ref{Dnm1})) is suppressed by the factor $%
e^{-2n\eta _{1}(\beta \sin \theta )}$. This may not be the case for the
Smith-Purcell part in the radiation intensity, coming from the last term in (%
\ref{Dnm1}). Indeed, for values of $m$ for which the order $n+mN+\alpha $ of
the Bessel function is not large by the modulus, one has $J_{n+mN+\alpha
}(n\beta ^{\prime }\sin \theta )\sim 1/\sqrt{n}$ and for the radius of the
circular orbit sufficiently close to the radius of the grating (in order to
escape the exponential suppression coming from the ratio of the Hankel
functions in (\ref{Dnm1})), under the condition $\eta (\beta ^{\prime }\sin
\theta )/\eta (\beta \sin \theta )-1\ll 1$, the Smith-Purcell part in the
radiation intensity dominates. By taking into account (\ref{etaclose}), this
condition can be written as%
\begin{equation}
1-\frac{r_{c}}{r_{e}}\ll \frac{\eta _{1}(\beta \sin \theta )}{\sqrt{1-\beta
^{2}\sin ^{2}\theta }}.  \label{CondDom}
\end{equation}%
For the ratio of the Hankel functions in (\ref{Dnm1}) one has%
\begin{equation}
\frac{H_{n+\alpha }(n\beta \sin \theta )}{H_{n+\alpha }(n\beta ^{\prime
}\sin \theta )}\approx \frac{r_{c}}{r_{e}}\frac{\left( 1-y^{\prime 2}\right)
^{1/4}}{\left( 1-y^{2}\right) ^{1/4}}\frac{1+\alpha \sqrt{1-y^{2}}}{1+\alpha
\sqrt{1-y^{\prime 2}}}e^{-n\left[ \eta _{1}(y^{\prime })-\eta _{1}(y)\right]
},  \label{Hankrat}
\end{equation}%
with $y=\beta \sin \theta $ and $y^{\prime }=\beta ^{\prime }\sin \theta $.
Under the condition $n\gg 1/\eta _{1}(\beta \sin \theta )$ the Smith-Purcell
contribution to the radiation intensity contains the factor (\ref{supfactor}%
). As a consequence, the values for the corresponding radiation harmonics
are restricted by the condition%
\begin{equation}
n\sqrt{1-\beta ^{2}\sin ^{2}\theta }(1-r_{c}/r_{e})\lesssim 1.  \label{Harmn}
\end{equation}%
Note that for $1-\beta \sin \theta \ll 1$ the condition (\ref{CondDom}) is
simplified to $1-r_{c}/r_{e}\ll 1-\beta \sin \theta $.

\subsection{Numerical illustrations}

For the further clarification of the radiation properties here we consider
numerical examples. Similar to the case of a metallic cylinder, we present
the results of the numerical investigations for the angular density of the
number of the radiated quanta per period of rotation,%
\begin{equation}
\frac{dN_{n}^{(g)}}{d\Omega }=\frac{T}{\hbar n\omega _{0}}\frac{dI_{n}^{(g)}%
}{d\Omega }.  \label{dNg}
\end{equation}%
\begin{figure}[tbph]
\begin{center}
\begin{tabular}{cc}
\epsfig{figure=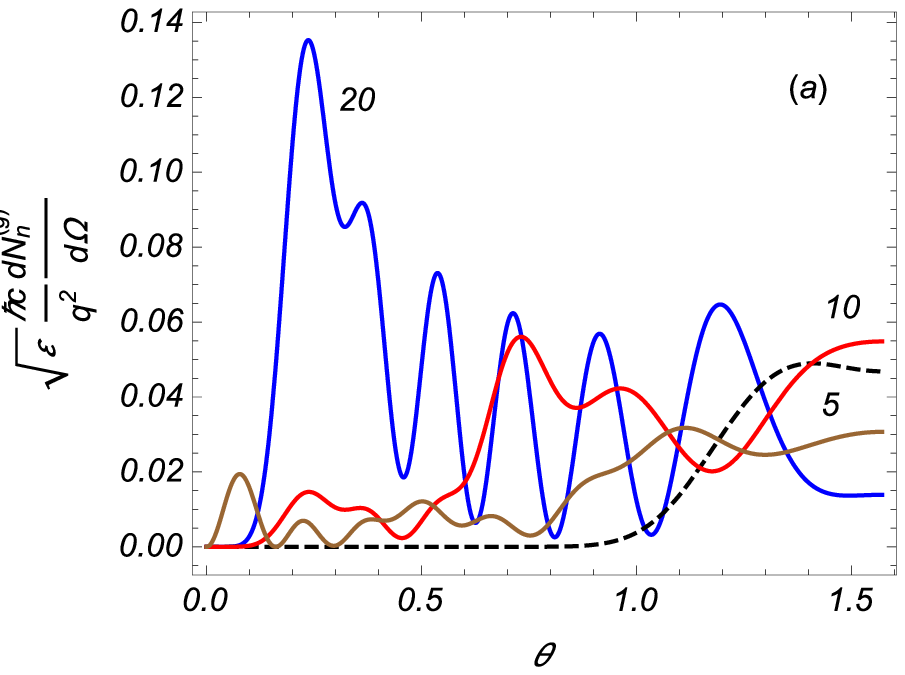,width=7.5cm,height=6.cm} & \quad %
\epsfig{figure=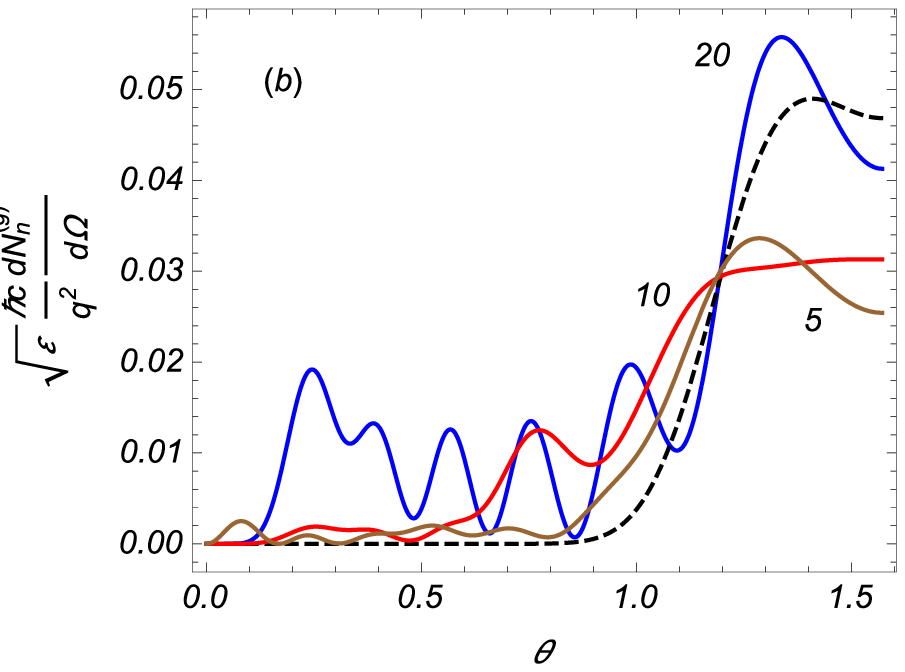,width=7.5cm,height=6.cm}%
\end{tabular}%
\end{center}
\caption{The angular dependence of the number of the radiated quanta per
period of the charge rotation for the electron energy 2 MeV and for the
radiation harmonic $n=25$. The numbers near the curves are the values of $N$%
. The panels (a) and (b) correspond to $r_{c}/r_{e}=0.99$ and $%
r_{c}/r_{e}=0.95$, respectively. For the other parameters we have taken $%
b/a=1$ and $\protect\varepsilon =1$.}
\label{fig5}
\end{figure}
In figure \ref{fig5}, for the electron with the energy 2 MeV, we have
plotted this quantity for the harmonic $n=25$ as a function of the angle $%
\theta $ (in radians) for different values of the parameter $N$ (the numbers
near the curves). The corresponding graphs in the region $\pi /2\leq \theta
\leq \pi $ are symmetric to the ones presented in figure \ref{fig5} with
respect to $\theta =\pi /2$. For the parameters of the grating we have taken
$b/a=1$ and $\varepsilon =1$. The panel (a) corresponds to $r_{c}/r_{e}=0.99$
and for the panel (b) $r_{c}/r_{e}=0.95$. The dashed curves correspond to
the radiation in the absence of the grating. As is seen, for the angles not
close to the rotation plane, the radiation intensity is dominated by the
Smith-Purcell part. For these angles one has the suppression of the
radiation with decreasing values of the ratio $r_{c}/r_{e}$. Note that $%
\sqrt{\varepsilon }dN_{n}^{(g)}/d\Omega $ depends on the parameters $%
\varepsilon $, $v$, $r_{e}$, $r_{c}$ in the form of the combinations $\beta $
and $r_{e}/r_{c}$. The parameters of the diffraction grating enter through
the factor $S_{m}$ from (\ref{Sms}) and through the number of periods $N$
appearing in the index of the Bessel function in (\ref{Dnm1}). Note that the
radiation wavelength is given by the relation $\lambda _{r}=2\pi
r_{e}/(n\beta )$.

\begin{figure}[tbph]
\begin{center}
\begin{tabular}{cc}
\epsfig{figure=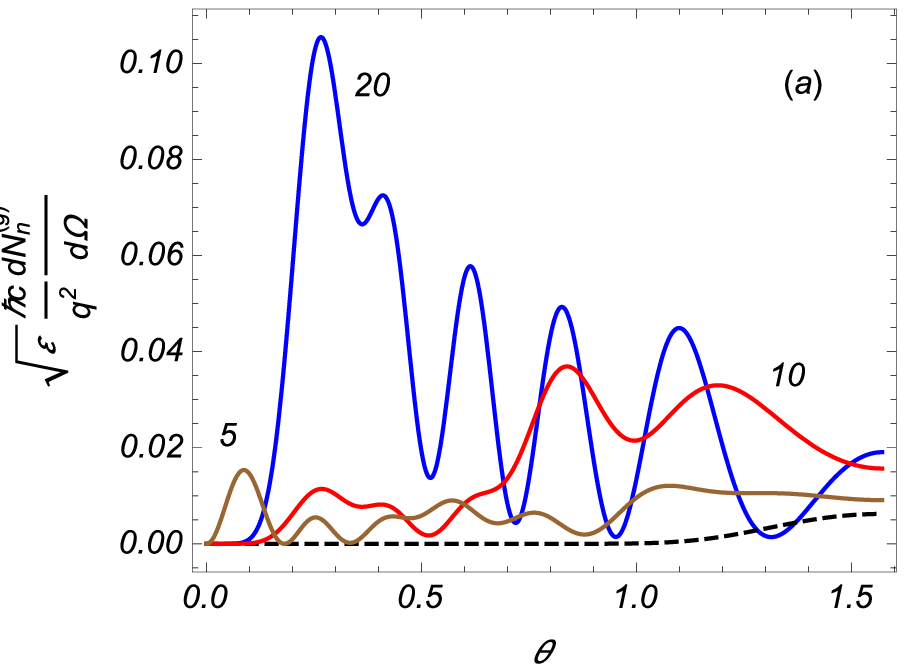,width=7.5cm,height=6.cm} & \quad %
\epsfig{figure=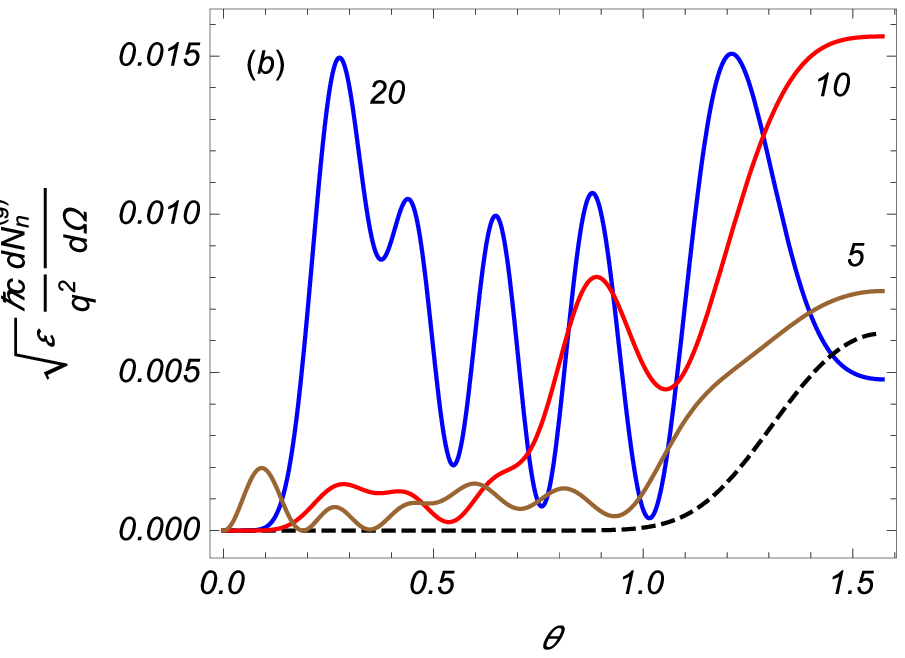,width=7.5cm,height=6.cm}%
\end{tabular}%
\end{center}
\caption{The same as in figure \protect\ref{fig5} for the energy of the
electron 1 MeV. }
\label{fig6}
\end{figure}

The same as in figure \ref{fig5}, for the energy of the electron 1 MeV, is
plotted in figure \ref{fig6}.
These graphs show that, with decreasing energy, the relative contribution of
the synchrotron radiation decreases and the Smith-Purcell part is dominant.
In accordance with the explanation given above, for large values of $n$ and $%
N$ the main contribution to the radiation intensity comes from the term with
the lowest value for $|n+mN+\alpha |$. In particular, for the examples with $%
n=25$, $N=20$ we have numerically checked that the dominant contribution in (%
\ref{dIg}) comes from the term in the series over $m$ with $m=-1$. The
locations of the local maxima and minima of the radiation intensity are
approximately determined by the local maxima and by the locations of the
zeros for the Bessel function $J_{n-N-1}^{2}(n\beta ^{\prime }\sin \theta )$.

\begin{figure}[tbph]
\begin{center}
\begin{tabular}{cc}
\epsfig{figure=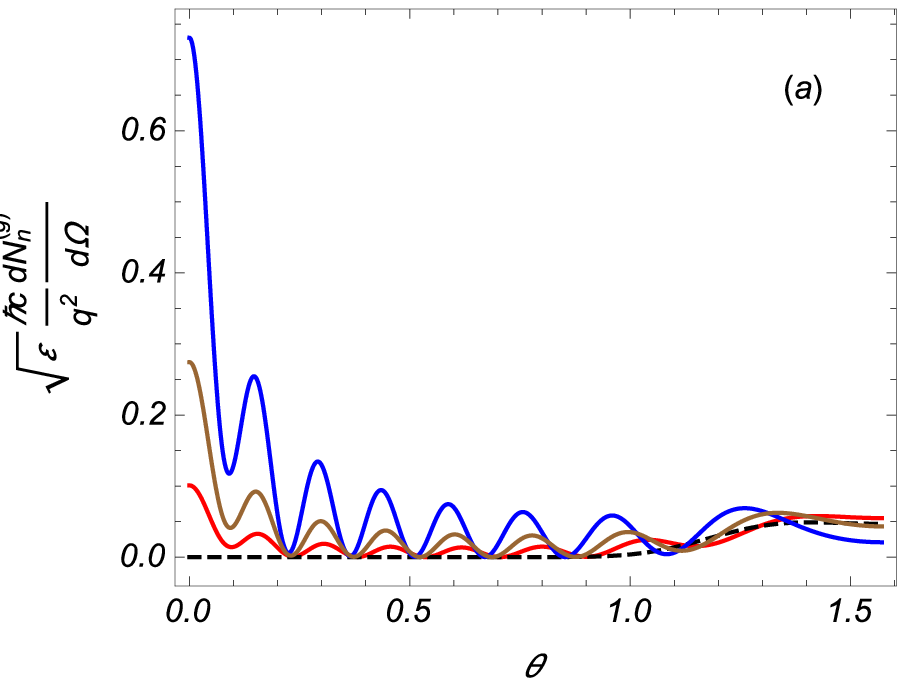,width=7.5cm,height=6.cm} & \quad %
\epsfig{figure=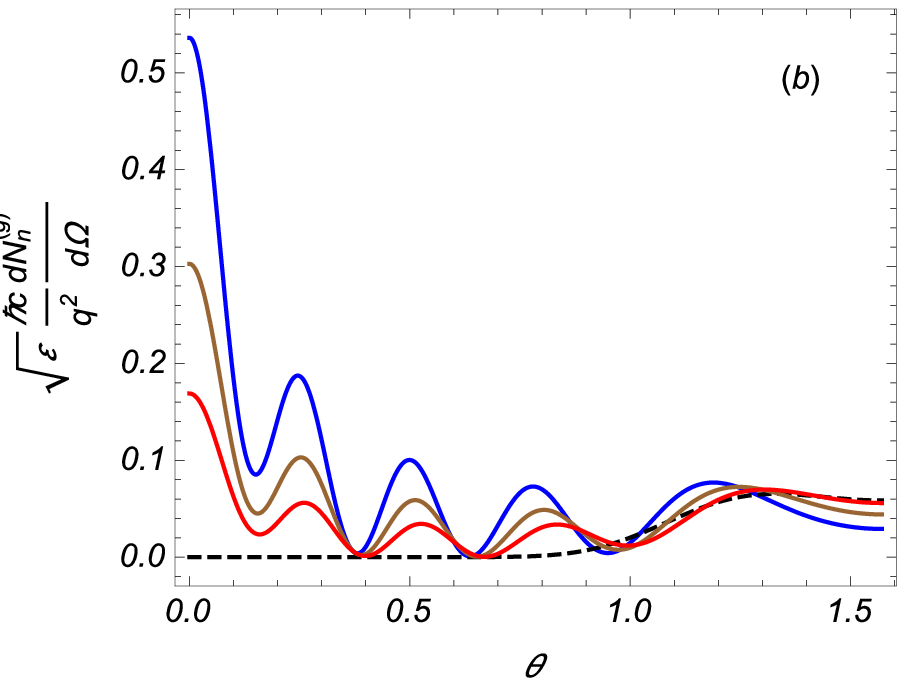,width=7.5cm,height=6.cm}%
\end{tabular}%
\end{center}
\caption{The angular density of the number of the radiated quanta as a
function of $\protect\theta $ for the electron energy 2 MeV and for $%
r_{c}/r_{e}=0.95$, $0.97$, $0.99$ (upper curves near $\protect\theta =0$
correspond to larger values of $r_{c}/r_{e}$). The panels (a) and (b) are
plotted for $n=25$, $N=24$ and $n=15$, $N=14$, respectively.}
\label{fig7}
\end{figure}

As it has been noted before, for a given harmonic $n$, the most strong
radiation at small angles $\theta $ is obtained for the number of periods $%
N=n\pm 1$. This feature is illustrated in figure \ref{fig7}, where the
angular dependence of the number of the radiated quanta is displayed for $%
r_{c}/r_{e}=0.95,0.97,0.99$. The curves with increasing values for $%
dN_{n}^{(g)}/d\Omega |_{\theta =0}$ correspond to increasing values of $%
r_{c}/r_{e}$. The panels (a) and (b) are plotted for $n=25$, $N=24$ and $%
n=15 $, $N=14$, respectively. The graphs for $N=26$ and $N=16$ have the
structure similar to the ones in panels (a) and (b), respectively. Again, we
have checked that for angles $\theta <1$rad the contribution of the term $%
m=-1$ is dominated in the radiation intensity and the oscillatory behavior
comes from the function $J_{n-N-1}^{2}(n\beta ^{\prime }\sin \theta )$. The
results depicted in figure \ref{fig7} show that, for given characteristics
of the charge (energy, radius of the orbit), by the choice of the parameters
for the diffraction grating, one can have highly directional radiation on a
given harmonic $n$ directed near the normal to the plane of the charge
rotation.

\begin{figure}[tbph]
\begin{center}
\begin{tabular}{cc}
\epsfig{figure=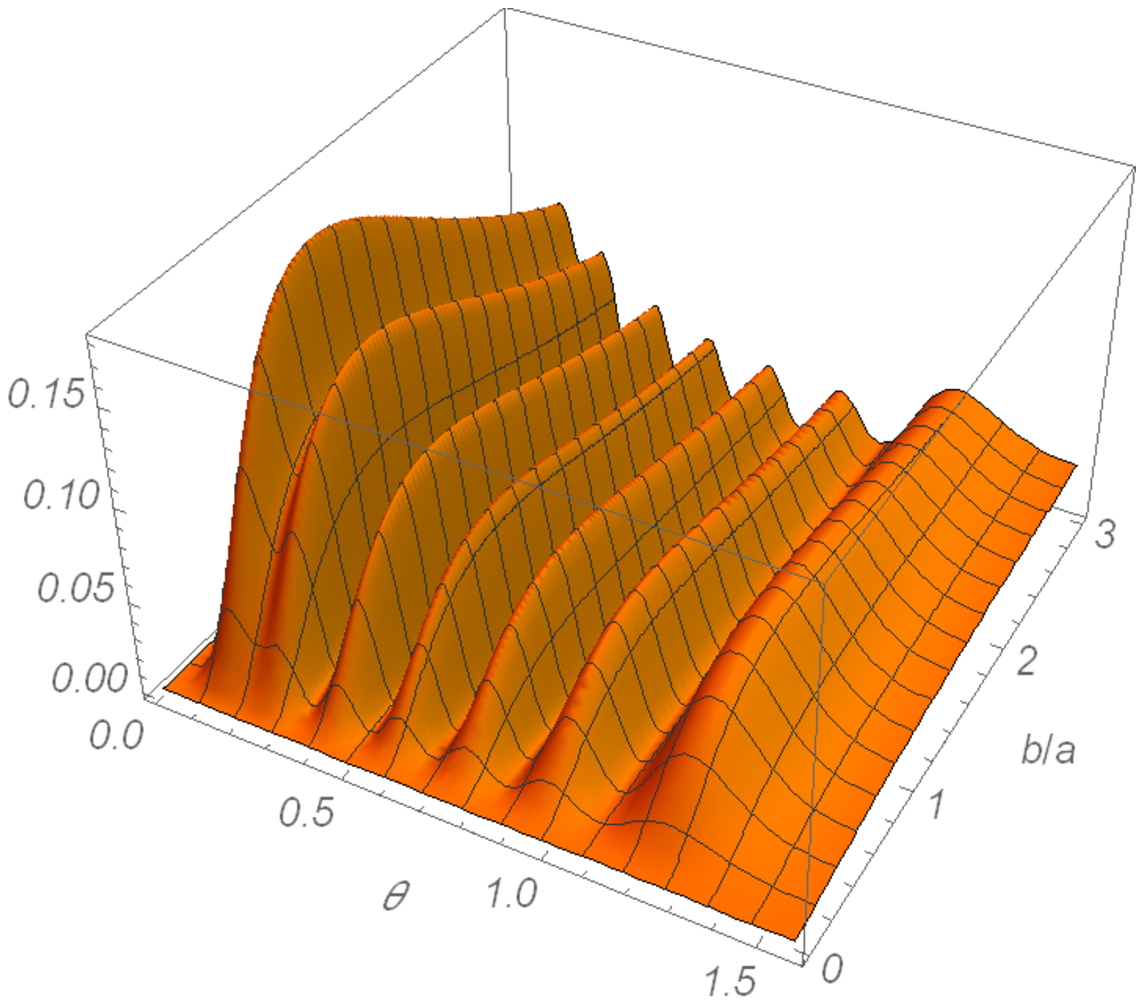,width=7.5cm,height=6.cm} & \quad %
\epsfig{figure=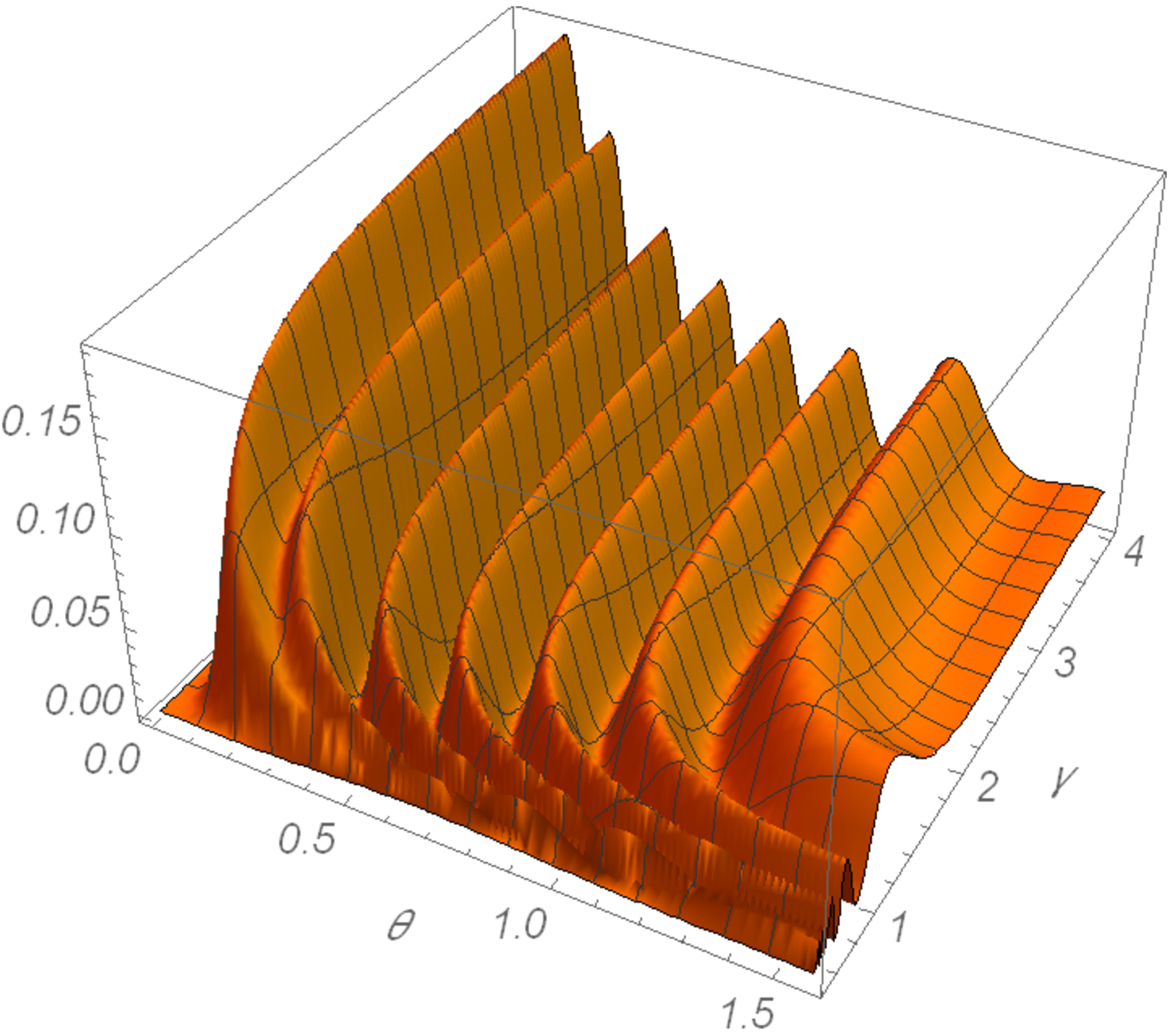,width=7.5cm,height=6.cm}%
\end{tabular}%
\end{center}
\caption{Number of the radiated quanta on the harmonic $n=25$ versus $%
\protect\theta $ and the ratio $b/a$ for the electron with the energy 2 MeV
(left panel) and versus $\protect\theta $ and the gamma factor for $b/a=1$
(right panel). For the values of the other parameters we have taken $N=28$, $%
r_{c}/r_{e}=0.99$, $\protect\varepsilon =1$.}
\label{fig8}
\end{figure}

It is also of interest to see the dependence of the radiation
characteristics on the ratio $b/a$ and on the energy of the radiating
charge. In figure \ref{fig8}, for the electron with the energy 2 MeV, we
present the angular density of the number of the radiated quanta on the
harmonic $n=25$ versus the angle $\theta $ and the parameter $b/a$ (left
panel) and versus $\theta $ and $\gamma =E_{e}/(m_{e}c^{2})$ for $b/a=1$
(right panel), with $E_{e}$ being the energy of the electron. For the other
parameters we have taken the values $N=28$, $r_{c}/r_{e}=0.99$, $\varepsilon
=1$. In the limit $b/a\ll 1$ we recover the result for a charge rotating
around a conducting cylinder. In the opposite limit $b/a\gg 1$ the result
for the rotation in the vacuum is obtained. As is seen, for the example
considered in figure \ref{fig8}, the locations of the angular peaks are not
sensitive to the ratio $b/a$. As a function of $b/a$, the radiation
intensity takes its maximum value for $b/a$ close to 1. From the right panel
of figure \ref{fig8} we see that, for $\gamma \gtrsim 3$ and for the angles $%
\theta $ not to close to $\pi /2$ (rotation plane), the heights and the
locations of the angular peaks in the radiation intensity are not sensitive
to the value of the electron energy $E_{e}$.

\begin{figure}[tbph]
\begin{center}
\begin{tabular}{cc}
\epsfig{figure=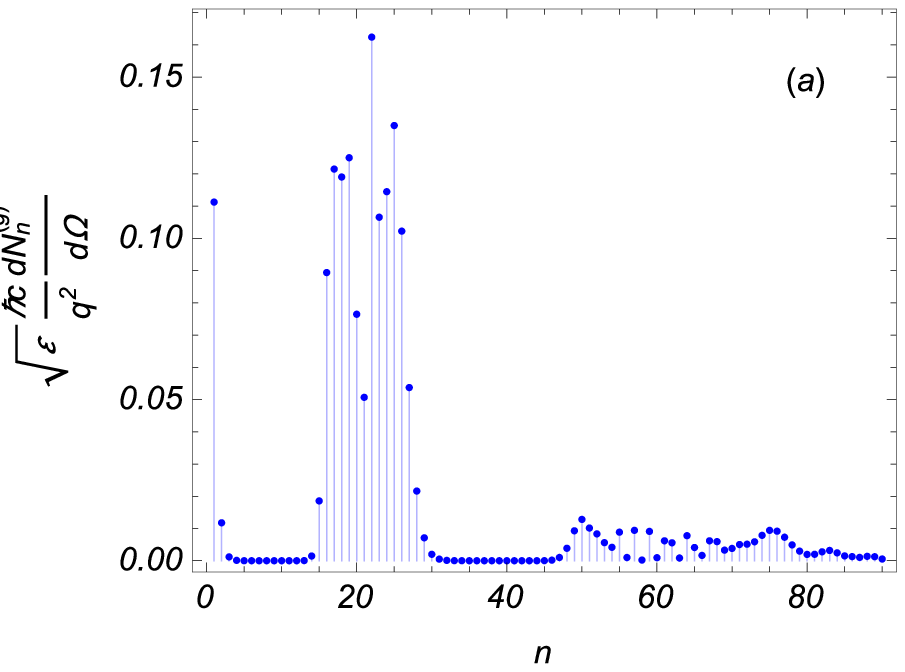,width=7.5cm,height=6.cm} & \quad %
\epsfig{figure=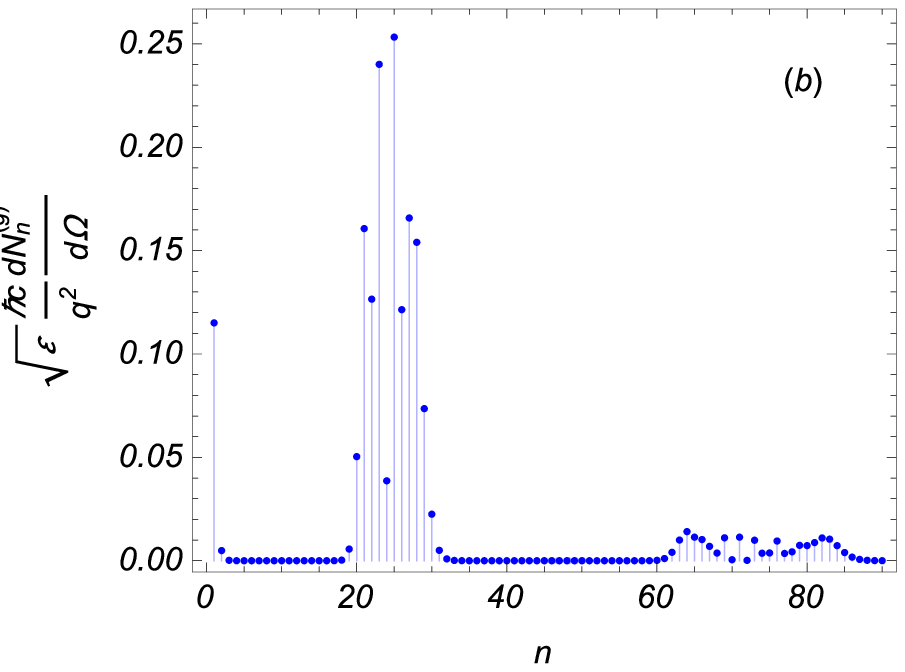,width=7.5cm,height=6.cm}%
\end{tabular}%
\end{center}
\caption{The angular density of the number of the radiated quanta as a
function of the radiation harmonic for the electron energy 2 MeV and for $%
r_{c}/r_{e}=0.99$. For the panels (a) and (b) one has $N=20$, $\protect%
\theta =0.24$ and $N=24$, $\protect\theta =0.15$, respectively.}
\label{fig9}
\end{figure}

In order to see the dependence on the number of the radiation harmonic, in
figure \ref{fig9} we have displayed the number of the radiated quanta as a
function of $n$ for the electron of the energy 2 MeV. The panels (a) and (b)
are plotted for $N=20$, $\theta =0.24$ and $N=24$, $\theta =0.15$. These
values of the angle correspond to the local peaks in the angular
distribution for the examples presented in figures \ref{fig7} and \ref{fig8}%
. For these angles the radiation is mainly dominated by the Smith-Purcell
part. The synchrotron part of the radiation is mainly located near the angle
$\theta =\pi /2$. The corresponding radiation intensity, $%
dI_{n}^{(0)}/d\Omega $, increases with increasing $n$ up to the values $n_{%
\mathrm{max}}\approx \gamma ^{3}$. For the energy corresponding to figure %
\ref{fig9} one has $n_{\mathrm{max}}\approx 60$. As regards the angular
density of the number of the radiated quanta, $dN_{n}^{(0)}/d\Omega $, it
monotonically decreases with increasing $n$.

\section{Summary}

\label{sec:Conc}

We have investigated the radiation from a charge rotating around conductors
with cylindrical symmetry. First the problem is considered with a charge
rotating around a conducting cylinder immersed in a homogeneous and
isotropic medium. The Fourier components of the electromagnetic fields,
defined in accordance with (\ref{Fl}), are decomposed into the parts
corresponding to a charge rotating in a homogeneous medium, in the absence
of the cylinder, and into the parts induced by the presence of the cylinder
(see (\ref{Fdec})). The Fourier components of the total fields are given by
the expressions (\ref{Hnz}) and (\ref{Enz}) with the coefficients defined in
(\ref{Bn}). The expressions for the cylinder induced parts are obtained from
(\ref{Hnz}) and (\ref{Enz}) replacing $B_{n+\alpha }$ by the second term in
the right-hand side of (\ref{Bn}). Having the fields, we have evaluated the
surface charge and current densities induced by the field of the charge on
the cylinder surface. For the Fourier component of the surface charge
density one has the expression (\ref{sign}). The only nonzero component of
the surface current density corresponds to the current along the azimuthal
direction and it is connected to the charge density by the standard
relation. We have explicitly shown that for the total surface charge one has
the relation (\ref{TotCharge}). Note that, for large values of the harmonic $%
n$, the part in the field induced by the cylinder is approximated by the
field of an effective point charge (\ref{qim}) located on the cylinder
surface. In the limit $r_{e}\rightarrow r_{c}$, the field of the charge is
compensated by its image and the fields in the region $r>r_{e}$ vanish.

For a charge rotating around a metallic cylinder, the angular density of the
radiation intensity on a given harmonic is given by the expression (\ref{dIn}%
). For the radius of the rotation orbit close to the cylinder surface, the
radiation intensity vanishes as $\left( r_{e}/r_{c}-1\right) ^{2}$. For
small angles $\theta $, one has the relation (\ref{SmallAng}) and the
radiation intensity behaves as $\sin ^{2(n-1)}\theta $. In the problem with
the conducting cylinder, the radiation intensity vanishes at zero angle for
all harmonics including $n=1$. For large harmonics of the radiation, under
the condition $\beta \sin \theta <1$, the effects in the radiation intensity
induced by the cylinder, compared with the standard synchrotron radiation,
are suppressed by an additional factor (\ref{supfactor}). For relativistic
particles and for radiation angles close to the rotation plane the
contribution of the cylinder to the total radiation intensity can be
essential for wavelengths much smaller than the distance from the cylinder.

In section \ref{sec:Grat}, we have studied the radiation for a charge
rotating around a metallic diffraction grating on a cylindrical surface. The
effect of the grating on the radiation intensity is approximated by the
surface currents induced on the strips by the field of the rotating charge.\
The electric and magnetic fields are presented in the decomposed form (\ref%
{Flt}) where the second term is the contribution induced by the grating. For
the latter one has the Fourier expansion (\ref{Fsl}) with the Fourier
components (\ref{Hs}) and (\ref{Es}). Compared with the case of the solid
cylinder, the Fourier expansion in the geometry of a grating contains an
additional summation over $m$. This corresponds to the periodic structure
along the azimuthal direction. At large distances from the grating, the
spectral-angular density of the radiation intensity is given by the formula (%
\ref{dIg}) with the functions $R_{n,m}^{(\alpha )}$ defined in accordance
with (\ref{Dnm1}). The second term in the right-hand side of (\ref{Dnm1})
determines the contribution of the Smith-Purcell radiation. In two limiting
cases $a\rightarrow 0$ and $b\rightarrow 0$, the expression (\ref{dIg})
coincides with the exact results for the radiation in a homogeneous medium
and for the radiation from a charge rotating around a solid cylinder. Unlike
to these limiting cases, the radiation intensity for the geometry of
diffraction grating on the harmonics $n>1$ does not vanish for small angles $%
\theta $. The corresponding leading term is given by (\ref{dIgsmtet}).

For a charge rotating around a diffraction grating, the behavior of the
radiation intensity on large harmonics $n$ can be essentially different from
that for a charge rotating in the vacuum or around a solid cylinder. For a
given radiation direction $\theta $, under the conditions $\beta \sin \theta
<1$ and $n\gg 1/\eta _{1}(\beta \sin \theta )$, the part in the radiation
intensity corresponding to the pure synchrotron radiation is exponentially
suppressed and, for the impact parameter obeying the condition (\ref{CondDom}%
), the total intensity is dominated by the Smith-Purcell part. These
features are illustrated in figures \ref{fig5}-\ref{fig7} where the
dependence of the angular density of the number of the radiated quanta is
depicted as a function of $\theta $ for various values of the parameters in
the problem. With decreasing energy, the relative contribution of the
synchrotron radiation decreases and the Smith-Purcell part is dominant. In
particular, the numerical analysis shows that, for given characteristics of
the charge, by the choice of the parameters of the diffraction grating, one
can have highly directional radiation on a given harmonic directed near the
normal to the plane of the charge rotation. For large values of the
radiation harmonic and of the number of periods in the diffraction grating,
the main contribution to the radiation intensity comes from the term in the
summation over $m$ with the lowest value for $|n+mN+\alpha |$ and the
locations of the angular peaks in the radiation intensity are not sensitive
to the values of the ratio $b/a$. In addition, for the angles $\theta $ not
to close to the rotation plane and for high-energy particles, the heights
and the locations of the angular peaks are not sensitive to the value of the
charge energy.

\end{document}